\pdfoutput=1
\documentclass[twocolumn]{aastex61}
\newcommand{\enzo}{{\tt Enzo}}

\submitjournal{Astrophysical Journal Letters}
\received{\today}

\shorttitle{Star Formation around a Massive Black Hole Seed}
\shortauthors{Aykutalp et al.}

\begin{document}

\title{Induced Metal-free Star Formation around a Massive Black Hole Seed} 

\correspondingauthor{Aycin Aykutalp}
\email{aycina@lanl.gov}

\author{Aycin Aykutalp}
\affiliation{Los Alamos National Laboratory, Los Alamos, New Mexico, 87545, USA}

\author{Kirk S. S. Barrow}
\affiliation{Kavli Institute for Particle Astrophysics and Cosmology, Stanford University, Stanford, California, USA}

\author[0000-0003-1173-8847]{John H. Wise}
\affiliation{Center for Relativistic Astrophysics, Georgia Institute of Technology, 837 State Street, Atlanta, GA 30332-0430, USA}

\author{Jarrett L. Johnson}
\affiliation{Los Alamos National Laboratory, Los Alamos, New Mexico, 87545, USA}

\begin{abstract}
The direct formation of a massive black hole is a potential seeding mechanism of the earliest observed supermassive black holes.  We investigate how the existence of a massive black hole seed impacts the ionization and thermal state of its pre-galactic host halo and subsequent star formation.  We show that its X-ray radiation ionizes and heats the medium, enhancing $\rm{H}_2$ formation in shielded regions, within the nuclear region in the span of a million years.  The enhanced molecular cooling triggers the formation of a $\sim 10^4~{\rm M}_\odot$ metal-free stellar cluster at a star formation efficiency of $\sim 0.1\%$ in a single event.  Star formation occurs near the edges of the \ion{H}{2} region that is partially ionized by X-rays, thus the initial size depends on the black hole properties and surrounding environment.  The simulated metal-free galaxy has an initial half-light radius of $\sim 10$~pc but expands to $\sim 50$~pc after 10 million years because of the outward velocities of their birth clouds.  Supernova feedback then quenches any further star formation for tens of millions of years, allowing the massive black hole to dominate the spectrum once the massive metal-free stars die.  
\end{abstract}


\section{Introduction} \label{sec:intro}
The increasing number of observations of quasars powered by supermassive black holes (SMBHs) at high redshifts ($z>6$) \citep{2006AJ....131.1203F, 2007ApJ...669...32K, 2011Natur.474..616M, 2015Natur.518..512W, 2017arXiv171201860B} continue to challenge black hole (BH) seeding and growth theories. Their seed BHs might form through (a) the growth of BH remnants from metal-free (Population III; Pop III) stars \citep{2003ApJ...582..559V, 2005ApJ...633..624V, 2007MNRAS.374.1557J}, (b) the collisions of stars in young dense stellar clusters \citep{1978MNRAS.185..847B, 2001ApJ...562L..19E}, or (c) the monolithic collapse of atomic cooling halos, so-called direct collapse black holes (DCBHs) \citep{1993MNRAS.263..168H, 1993ApJ...419..459U, 2003ApJ...596...34B, 2006ApJ...652..902S, 2008ApJ...682..745W}. Independent of the formation scenarios listed above, the current accretion prescriptions cannot easily explain the rapid growth of these seeds into SMBHs within 900 Myr, especially when the radiative feedback effects from the accreting BH are taken into account \citep{2011ApJ...738...54K,  2013MNRAS.432.3438A, 2014ApJ...797..139A}.  However, having a more massive seed as the precursor of these SMBHs, such as a DCBH, has some initial merit, and thus it has been the focus of many recent cosmological studies.

In order to form a DCBH in the early universe, there are certain conditions that need to be met so that collapsing primordial gas will not fragment into smaller clumps and subsequently forming stars. Current DCBH formation prescriptions in numerical simulations mostly include Lyman Werner (LW; $E = 11.2-13.6 \, \rm{eV}$) background radiation, produced by a stellar population in a nearby halo, that dissociates H$_2$ in the collapsing cloud. Thereby, suppressing radiative cooling and fragmentation into smaller clumps. Simulations on the photodissociation of H$_2$ in protogalaxies have shown that a LW flux in excess of $\sim 20~J_{21}$ ($2000~J_{21}$) originating from a $10^4$~K ($10^5$~K) blackbody spectrum is sufficient to prevent H$_2$ formation and, hence, fragmentation in halos with virial temperatures of $T_{\rm vir} \sim 10^4$~K \citep[e.g.][]{2010MNRAS.402.1249S, 2014MNRAS.443.1979L}.  Here $J_{21}$ is the specific intensity just below $13.6 \, \rm{eV}$ ($J_{21} = 10^{-21} \, \rm{erg} \, \rm{s}^{-1} \, \rm{cm}^{-2} \, \rm{sr}^{-1} \, \rm{Hz}^{-1}$). However, even though the nominal critical $J_{21}$ flux is much lower in the $10^4$~K case, it requires more mass in stars to suppress H$_2$ cooling compared to $10^5$~K case as studied in \citet{2012MNRAS.425L..51W}. \citet{2012AIPC.1480..309I} and \citet{ 2014MNRAS.442L.100V} argued that there is a zone of no return for a collapsing gas cloud to form a DCBH without fragmenting into smaller clumps, which depends on the density (${n} > 10^4 \, \rm{cm^{-3}}$) and temperature (${T}\geq10^4$ K) of the collapsing gas. More recently, it has shown that choosing one fixed value for the LW flux causes under- or over-estimation of DCBH formation sites and that one needs to consider the precise age, mass or star formation rate, distance of the stellar population producing the radiation field \citep{2016MNRAS.459.4209A}. \citet{2017A&A...601A.138J} show that when the detachment of H$^-$ by Lyman $\alpha$ photons is considered, the required LW radiation intensity to suppress efficient H$_2$ formation decreases by $1-2$ orders of magnitude in the case of a $10^5 \, ~ \rm{K}$ blackbody spectrum. In their recent work, \cite{2020arXiv200105498W} also found that the H$^-$ photo-detachment by Lyman $\alpha$ photon decreases the critical UV flux by up to a factor of $\sim 5$. Furthermore, \citet{Wise_2019} have shown that dynamical heating driven by major mergers can further induce DCBH formation in overdense regions. 

In this Letter, we investigate the early evolution of a DCBH hosting halo. This halo initially starts as star-less but with a central dense gaseous core and central BH.  However, its state rapidly changes once the accretion radiation from the DCBH interacts with its environment.  The X-ray irradiation from the accreting DCBH ionizes the ambient gas which has a profound effect on the subsequent evolution, triggering metal-free star formation and resulting in an ``obese BH galaxy'' \citep[OBG;][]{2013MNRAS.432.3438A}.  Here, we quantify the effects of X-ray irradiation from the central accreting DCBH on the formation of the stellar population and shaping the evolution of the host halo. We studied the observability prospects of such a system at high redshift and established that it is possible to distinguish halos with and without DCBHs at their centers with James Webb Space Telescope (JWST) in \citet{Barrow_2018}. In our simulations, LW radiation from the accreting DCBH is not taken into account (see Section \ref{sec:dis} for detailed discussion).

This letter is structured as follows. In Section 2, we describe our methods. We present our results and implications in Section 3. Finally in Section 4, we discuss and summarize our conclusions.

\section{Methods} \label{sec:meths}
We further analyze a cosmological simulation from \citet{2014ApJ...797..139A} that is performed with the Eulerian adaptive mesh refinement (AMR) hydrodynamic code \enzo{} \citep{2014ApJS..211...19B}. 
The simulation has a $(3\, \rm{ Mpc})^3$ comoving box size with a $128^3$ root grid resolution and three static nested grids, each refined by a factor of two with the innermost grid having an effective resolution of 1024$^3$ with a side length of 375 h$^{-1}$ kpc. During the course of the simulation, we allow a maximum refinement level $l=10$, resulting in a maximal resolution of 3.6 proper parsecs. Refinement is restricted to the finest nested grid and occurs on baryon and dark matter overdensities of $3 \times 2^{-0.2l}$. Here $l$ is the AMR level, and the negative exponent means that the mass resolution in the calculations is super-Lagrangian. The simulation is initialized at $z=99$ by utilizing {\tt inits}, a package that uses Zel'dovich approximation and use the Wilkinson Microwave Anisotropy Probe seven-year cosmological parameters $\Omega_{\Lambda} = 0.734$, $\Omega_{\rm m} = 0.266$,  $\Omega_{\rm b} = 0.0449$, $\sigma_8 = 0.81$, and $h=0.701$ \citep{2011ApJS..192...18K} with standard definitions for each variable.

To provide the conditions for the DCBH to form at the center of a halo, we consider a strong, uniform, time-independent LW radiation background of $10^3 \, J_{21}$, emulating a nearby young galaxy irradiating this primordial atomic cooling halo with a virial mass $M_{\rm vir} = 2.6 \times 10^8 \, M_{\odot}$ at $z=15$.  Prior to seeding the DCBH of mass $5 \times 10^4~M_{\odot}$ at $z=15$ there was no prior star formation in the entire simulation volume.

\subsection{XDR Grid \& Treatment of the Polychromatic Spectrum} \label{sec:xdr}

We utilize a modified version of the X-ray and photon dominated region (XDR/PDR) code by \citet{2005A&A...436..397M} to compute the pre-calculated grids for temperatures and species abundances for a given X-ray flux ($F_{\rm X}$), hydrogen density ($n_{\rm HI}$), hydrogen column density ($N_{\rm HI}$), and metallicity ($Z/Z_{\odot}$). We include all the heavy elements up to iron with abundances $>10^{-6}$ relative to hydrogen as well as the singly and doubly ionized states of all elements, including He$^{2+}$. Secondary ionization by energetic electrons produced from primary X-ray ionization are more important for H, H$_2$, and He than primary ionizations. In our treatment, we took into account its effect on the thermal and chemical properties of the gas, see \citet{2013ApJ...771...50A} and references therein for further details. We modified the \enzo{} code to use this grid where the thermodynamical evolution of the gas is driven by the XDR physics. We use \enzo{'s} nine-species (H, H$^+$, H$^-$, He, He$^+$, He$^{2+}$, H$_2$, H$^-_2$, and e$-$) non-equilibrium chemical network for zero metallicity case \citep{1997NewA....2..181A, 1997NewA....2..209A}. When there is an impinging X-ray radiation onto a cell, we compare the temperatures calculated from the XDR grid and the \enzo{'s} chemical network and take the highest value of the two found temperatures and continue to iterate for the next step. By taking the highest, we divide the simulation box into XDR and non-XDR zones. We might overestimate the temperature in the grid when X-ray heating and non-X-ray heating are comparable, but this pertains to a very small part of the grid given the deep penetration of X-rays into dense gas. 

The radiative feedback from the accreting DCBH is treated by using the radiation transport module {\tt Moray} \citep{Wise11_Moray}. We use an $N_{\rm HI}$ lookup-table for a polychromatic X-ray spectrum to calculate the attenuation in each line of sight \citep{2006NewA...11..374M, 2014ApJ...797..139A}. To construct the table, the radiative transfer equation is numerically solved before the simulation, giving a relative ionizing photon flux $I_\nu$ as a function of the $N_{\rm HI}$. The relative ionizing photon flux for \ion{H}{1}, \ion{He}{1}, and \ion{He}{2} is computed and stored for 300 column densities, equally log-spaced over the range $N_{\rm HI} = 10^{12}-10^{25} \,\rm{cm}^{-2}$. The details of this approach are described in \citet{2013ApJ...771...50A}.

\subsection{Accretion Prescription}\label{sec:acc}

For the accretion onto the DCBH we follow the prescription discussed in \citet{2011ApJ...738...54K} where accretion rate is estimated according to the Eddington-limited spherical Bondi-Hoyle \citep{1952MNRAS.112..195B} formula 

\begin{eqnarray}
\dot{M}_{\rm{BH}} &=& \mathrm{min} (\dot{M}_{\rm{B}}, \dot{M}_{\rm{Edd}})\\
&=&\mathrm{min} \left(\frac{4\pi G^2 M_{\rm{BH}}^2\rho_{\rm{B}}}{c_{\rm s}^3}, \frac{4\pi G M_{\rm{BH}}m_{\rm{p}}}{\epsilon \sigma_{\rm{T}}c}\right). \nonumber
\end{eqnarray}
Here $G$ is the gravitational constant, $M_{\rm BH}$ is the DCBH mass, $c_{\rm s}$ is the sound speed, $m_{\rm p}$ is the proton mass, $\epsilon$ is the radiative efficiency, $\sigma_{\rm T}$ is the Thomson scattering cross-section, and $\rho_{\rm B}$ is the density at the Bondi radius $R_{\rm B} = 2 G M_{\rm BH}/ c_{\rm s}^2$. 

\subsection{Star Formation \& Feedback}\label{sec:sf}
The star formation in our simulations is only allowed in the finest AMR levels. A Pop III star particle forms when the molecular hydrogen fraction $f_{\rm H_2} > 5 \times 10^{-4}$, the metallicity of the gas $Z/Z_{\odot} < 10^{-3.5}$ and, the velocity flow is converging; i.e., $ \nabla \cdot v < 0$, as discussed in \citep{2007ApJ...659L..87A, 2008ApJ...685...40W, 2012ApJ...745...50W}. In our simulations, a Pop III star particle represents a single star, whereas a Pop II star particle represents a star cluster. The transition from Pop III to Pop II star formation occurs in the simulation when the metallicity of the collapsing gas becomes $Z/Z_{\odot} > 10^{-3.5}$. However, during the course of our simulations no Pop II star formation occurs. The initial mass of Pop III stars  are randomly sampled from an IMF with a functional form
\begin{equation}
f(\log M) dM = M^{-1.3} \exp\left[-\left(\frac{M_{\rm char}}{M}\right)^{1.6} \right] dM,
\end{equation}
where $M_{\rm char} = 40~M_\odot$ is the characteristic mass of the Pop III stars. For  H$_2$ self-shielding, which is crucial for star formation to occur in regions where H$_2$ column densities exceed $10^{14} \, \rm{cm^{-2}}$, we use the prescription given by \citet{1996ApJ...468..269D} and \citet{2010MNRAS.402.1249S} (see section \ref{sec:dis} for further discussion). 

\section{Results}\label{RES}

\begin{figure}[t]
\begin{center}
\includegraphics[angle=0,width=8.5cm]{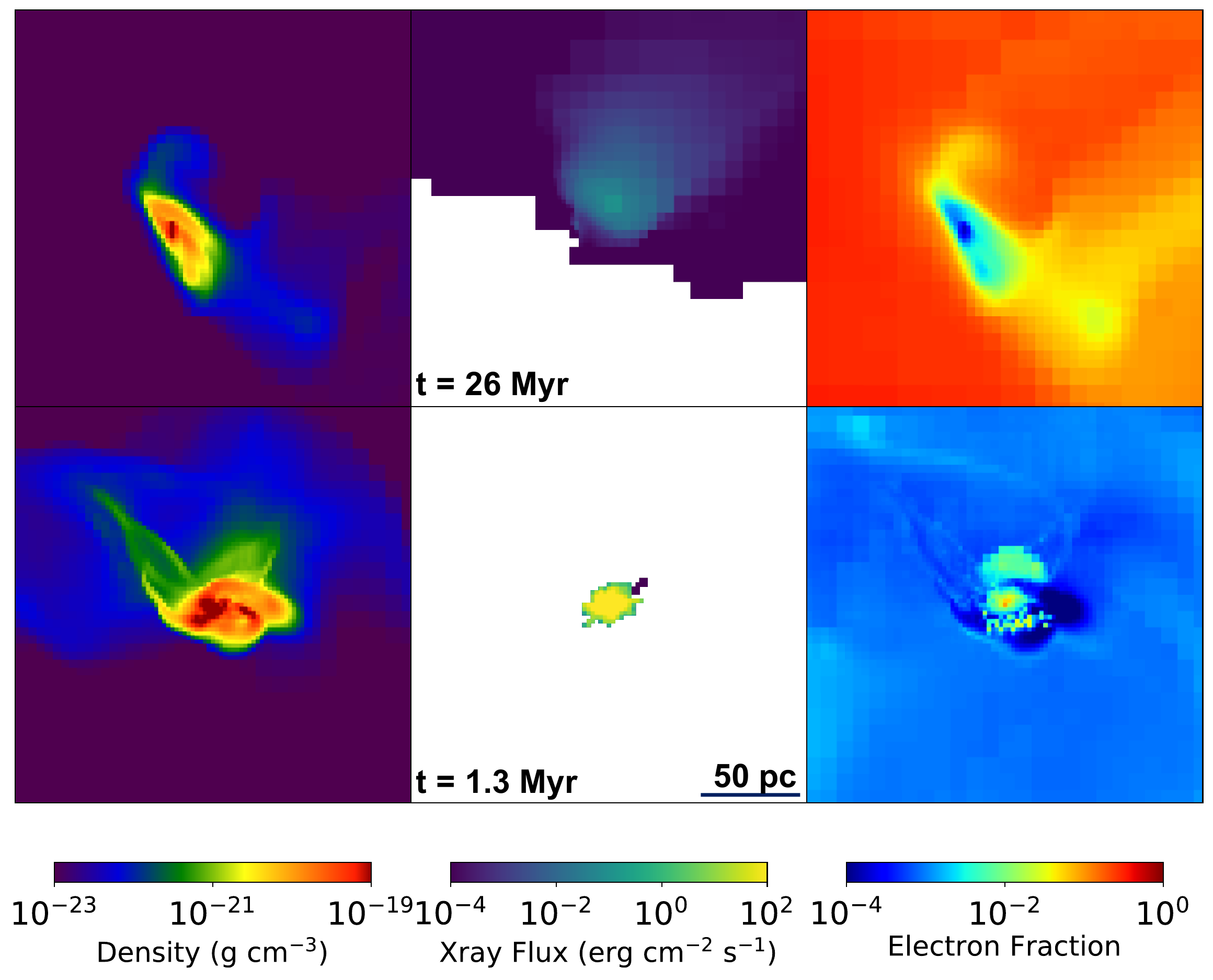}
\caption{Two-dimensional slices of the inner 200 parsecs of the gas density, X-ray flux and electron fraction at $ \rm t = 1.3\, \rm Myr$ (bottom) and  $ \rm t = 26 \, \rm Myr$ (top) after the formation of the DCBH.}\label{fig:F1}
\end{center}
\end{figure}

This work focuses on the consequences of DCBH presence within the host halo and, in particular, the initial star formation event. We show below that the newly formed DCBH alters the hydrodynamical state of the halo through its accretion radiation.  Figure \ref{fig:F1} shows two-dimensional slices of gas density (left), X-ray flux (middle), and electron fraction (right) of the inner 200 parsecs, 1.3 and 26  Myr after the DCBH formation, bottom and top, respectively. X-ray irradiation from the accreting DCBH ionizes the adjacent medium, and its large mean free path causes an extended transition from ionized to neutral. It is in these partially ionized regions where the additional free electrons facilitates H$^-$ and thus H$_2$ formation \citep{2001ApJ...560..580R}. 

\begin{figure}[t]
\hspace{-0.4 cm}
\includegraphics[angle=0,width=7.5cm]{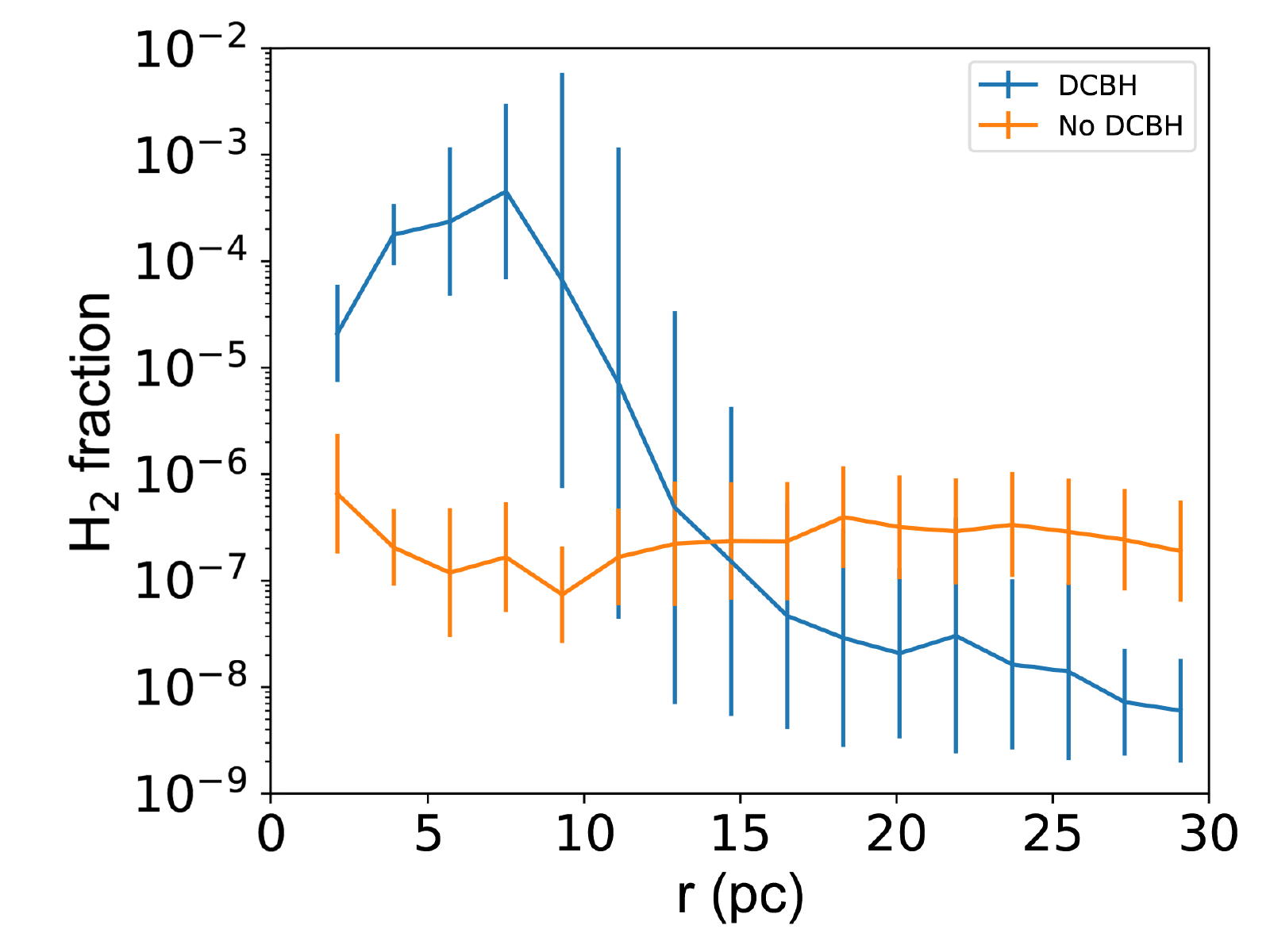}
\caption {Comparison of H$_2$ profile with (blue) and without (orange) the DCBH, $1.3\, \rm{ Myr}$ after the DCBH is formed. The error bars represent the standard deviation.}\label{fig:F2} 
\end{figure}

In Figure \ref{fig:F2}, we show the H$_2$ profile of the halo with (blue) and without (orange) the seed DCBH. The H$_2$ fraction of the ambient gas in the non-DCBH host halo stays around $\sim 10^{-7}$ where as the H$_2$ fraction in the DCBH host halo is boosted to $\sim 10^{-4}$ due to the X-ray radiation from the accreting DCBH. X-rays are completely attenuated in the inner 15 parsecs by the high density ambient gas (see the bottom-middle panel in Figure \ref{fig:F1}). Therefore, we only see the boost in the H$_2$ abundance within 15 parsecs of the DCBH at $t = 1.3$~Myr.

\begin{figure}[t]
\includegraphics[angle=0,width=8cm]{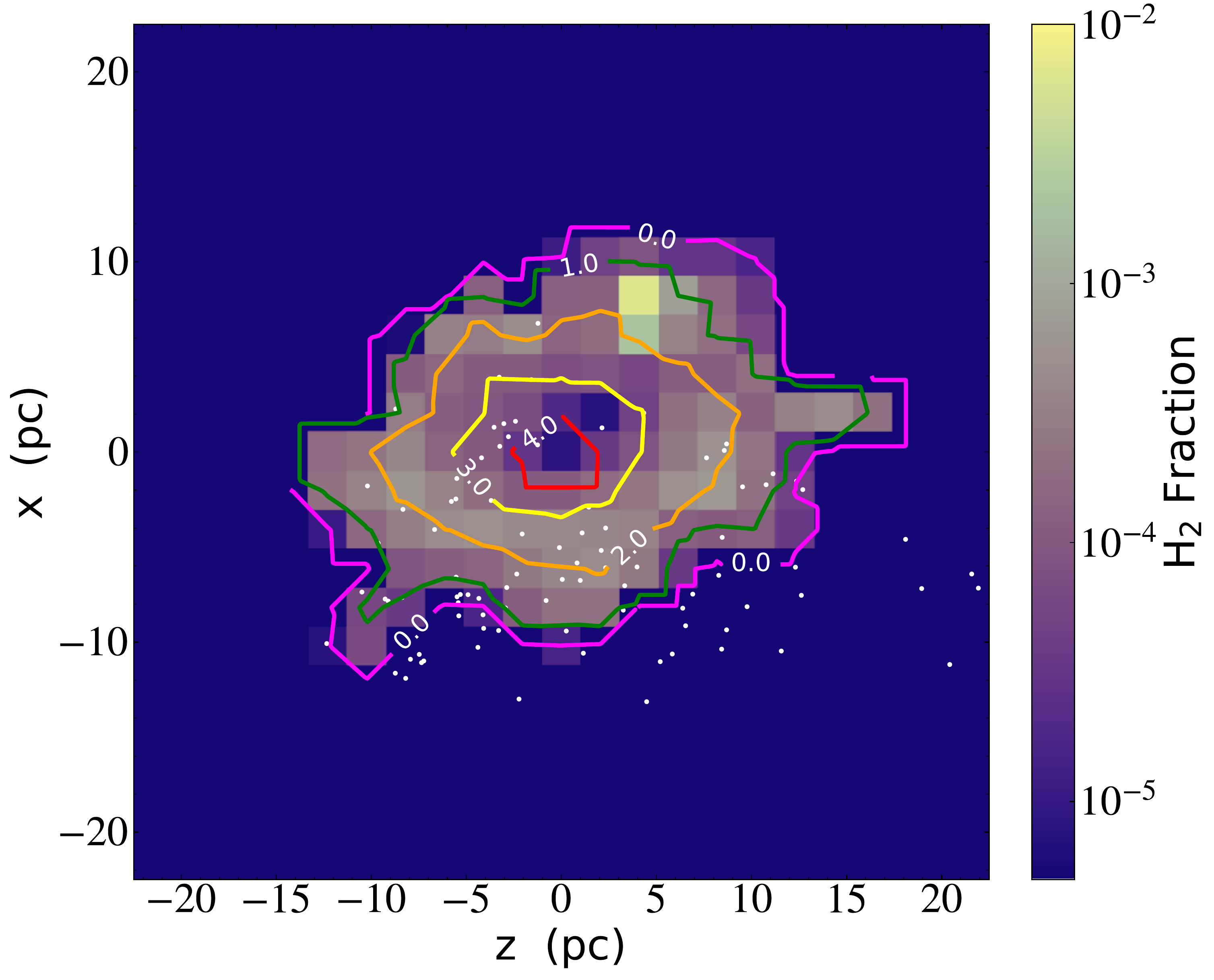}
\caption {A density-weighted projection of molecular hydrogen H$_2$ fraction at a time 1.3~Myr after the central DCBH starts accreting.  The contours show the X-ray flux in units of $\rm erg \,  cm^{-2} \, s^{-1}$ in log-scale. The white dots represent the Pop III stars that are induced by the X-ray irradiation.}\label{fig:F3}
\end{figure}

Despite the strong LW background radiation, the DCBH irradiates the nearby gas with X-rays, enhancing its H$_2$ formation and lowering the cooling time below the local dynamical time in particular regions at a distance $\sim$10 parsecs from the DCBH.  This cooling instability allows for metal-free (Pop III) star formation in the host halo in its nuclear region only $\sim 1$~Myr after DCBH formation.  Figure \ref{fig:F3} shows the density-weighted projection of H$_2$ fraction over plotted with X-ray contours. The white dots represent single Pop III star particles that are induced by the X-ray irradiation. There is a tight correlation between the path of X-ray radiation, enhancement of the H$_2$ fraction and the birth places of Pop III stars. Furthermore, the additional LW radiation from newly born Pop III stars keep H$_2$ fraction in the DCBH case at $\rm r > 15 \, \rm{pc}$ lower compared to non-DCBH case.

In the simulation, a total of 90 Pop III star particles formed, totaling 6932 $M_{\odot}$, equivalent to a star formation efficiency of $\sim$0.1\% within the central birth cloud.  It is difficult to predict an exact efficiency because of the uncertainties in the Pop III initial mass function (IMF) and the effects of the ensuing feedback.  However, it is clear that star formation will be triggered by the newly-formed DCBH, regardless of the DCBH mass or Pop III IMF, and both will alter the early chemo-thermal state of the galaxy.  

\begin{figure}[t]
\begin{center}
\includegraphics[angle=0,width=8cm]{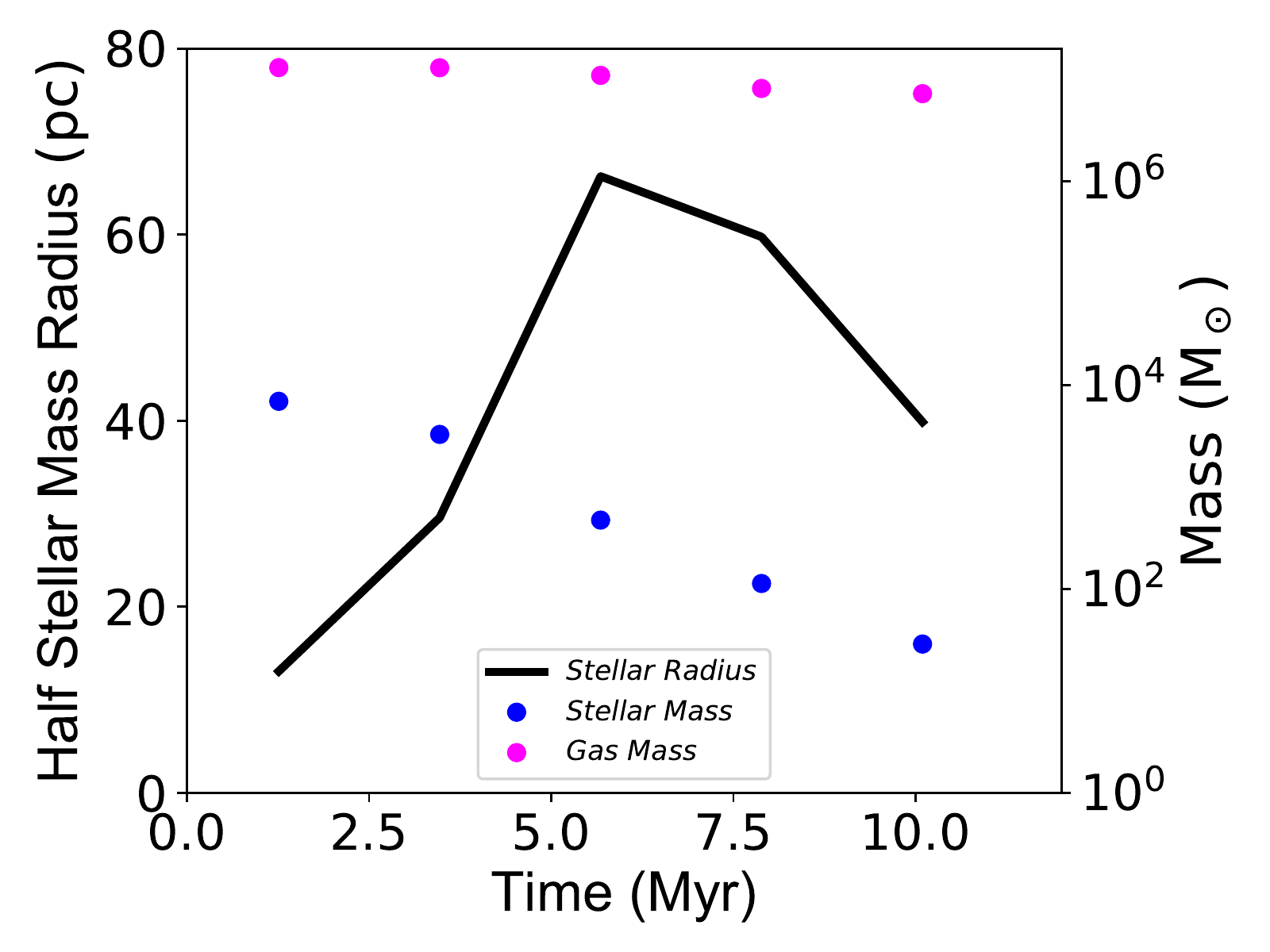}
\caption{Time evolution of the half stellar mass radius (solid black, left y-axis), total stellar mass (blue stars, right y-axis) and the gas mass (purple dots, right y-axis) within the half-mass radius. }
\label{fig:F4}
\end{center}
\end{figure}

The first star formation episode the DCBH host halo experiences is initially very centrally concentrated with the half stellar mass radius is being 12 parsecs as shown in Figure \ref{fig:F4}, where we plot the time evolution of half stellar mass radius (left y-axis), stellar mass (blue stars, right y-axis) and the gas mass (purple dots, right y-axis). This is in a very good agreement with the region that is enhanced by H$_2$, as shown in Figure \ref{fig:F2}. The first stars formed $\sim 1$ Myr after the formation of the DCBH. Due to energy injection by the SNe to the medium and the outward velocities of their birth clouds, the stars disperse to larger radii after they born as seen in Figures \ref{fig:F3} and  \ref{fig:F4}.

\begin{figure}[t]
\begin{center}
\includegraphics[angle=0,width=8cm]{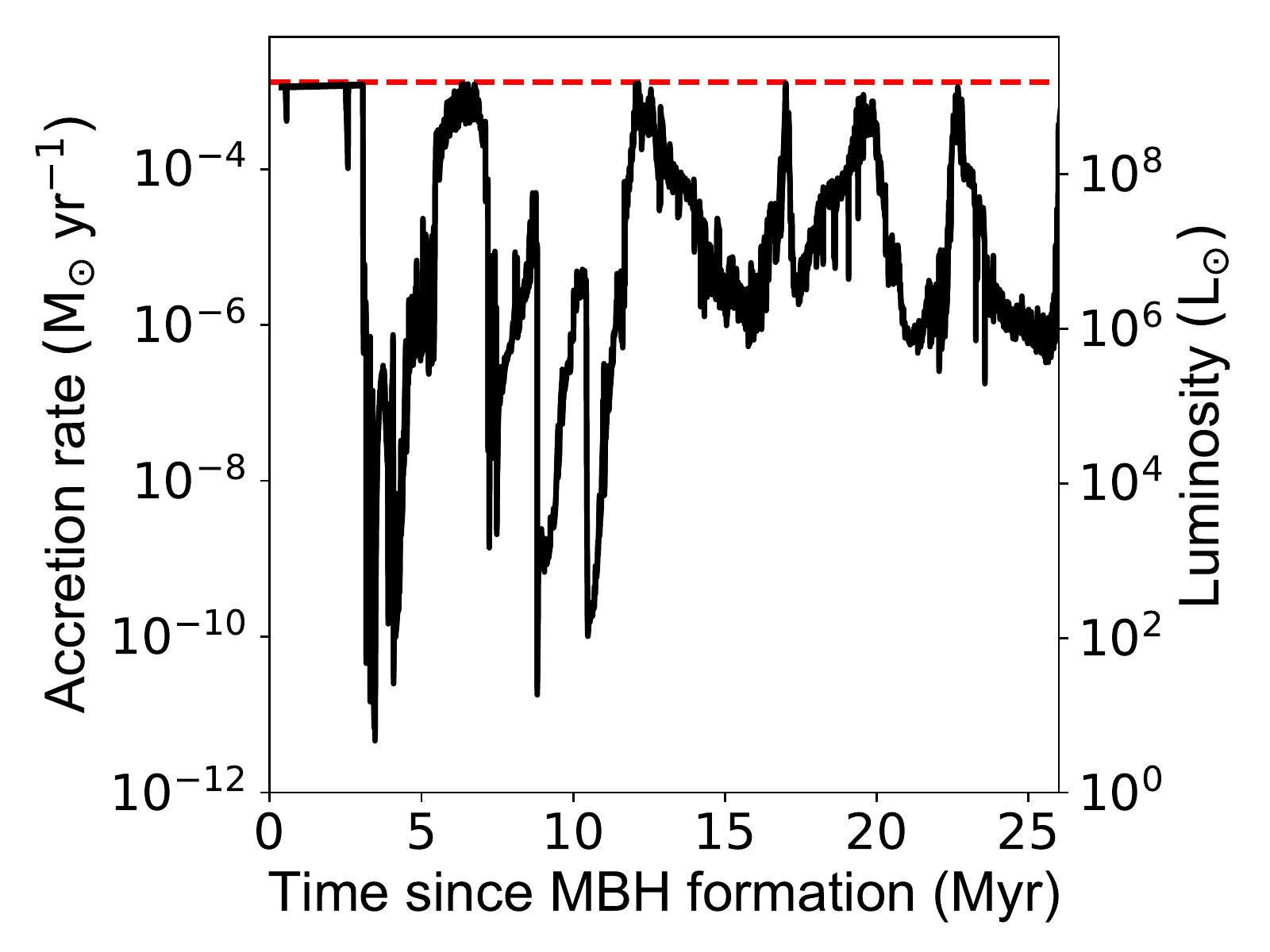}
\caption{Accretion rate (left y-axis) and corresponding luminosity (right y-axis) of the central DCBH over time. The red dashed line represents the Eddington limit.}
\label{fig:F5}
\end{center}
\end{figure}

Pop III stars in our simulation have short lifetimes $<12~\rm{Myr}$ due to their large initial masses, with $M_{\rm char} = 40~{M}_\odot$. At the end of their lifetime, a fraction of them explode as supernovae (SNe) and chemically enrich their surroundings. The binding energy of the inner 100 parsecs of the halo is on the order of $10^{53} \, \rm {ergs}$ whereas the energy released by the SNe is $10^{51} \, \rm {ergs}$. Hence, the gas mass within 100 parsecs does not decrease \citep{2005ApJ...630..675K}, see Figure \ref{fig:F4}. Throughout the SNe period the DCBH experiences lower accretion rates ($\sim 10^{-6}\, M_\odot \, \rm yr^{-1}$)  as shown in Figure \ref{fig:F5}. After all of the Pop III stars die, the DCBH starts accreting at high rates again ($\sim 10^{-3} \, M_\odot \, \rm yr^{-1}$) and produces X-ray. But now the ambient gas is enriched by the metals which has a vital affect on the temperature of the gas. The X-rays above 1~keV have small mean free paths in metal-enriched gas and are absorbed by inner shell electrons of C, N and O. Due to the high heating efficiency of X-rays the gas temperatures are kept around $10^5-10^6\, \rm{K}$. As shown in the top panel of Figure \ref{fig:F1}, X-rays penetrate to large distances ($\rm r > 200$ pc) and ionize the medium. Thus, even though the ambient gas is enriched by the metals the halo does not experience Pop II formation for the duration of our simulations, $\sim 50\, \rm{Myr}$. 
\section{Summary \& Discussion} \label{sec:dis}
In this work, we show, for the first time, how a DCBH induces a metal-free nuclear starburst in the host halo. The DCBH accretion produces X-ray radiation that ionizes the dense ambient gas. This increase in the free electron abundance facilitates H$^-$ and hence H$_2$ formation within 15 pc of the DCBH.  The radiation enhances cooling rates by boosting H$_2$ abundances, that then triggers a gravitational instability and subsequent star formation.  Because one of the requirements for DCBH formation is metal-free gas, the following star formation event is mostly likely to be metal-free since the DCBH progenitor, i.e. a supermassive star, ejects little to no heavy elements during its evolution, and the vast majority of supermassive stars do not produce a SN \citep{10.1111/j.1365-2966.2010.17359.x, 2017ApJ...842L...6W}. We note that, in our simulations, we did not take into account the LW radiation from the accreting DCBH. A back of the envelope calculation shows that the strength of LW radiation, $J_{LW}$ from a $5\times10^4\ M_{\odot}$ DCBH accreting at Eddington rate ($10^{-3}\, M_{\odot}\, yr^{-1}$) would be $7 \times 10^4$  and $1.8  \times 10^6  \, J_{21}$ at 50 and 10 pc, respectively. We performed the same simulation for LW background of $10^5 J_{21}$ where star formation with a total stellar mass of 7128 $M_{\odot}$ occurs. Thus, we conclude that the overall character of star formation at the center of the halo is not strongly sensitive to the precise level of LW when $J\rm_{21} > 10^3$.

Our DCBH scenario resembles OBGs studied in \citep{2013MNRAS.432.3438A}, where the mass of the DCBH ($\rm M_{\rm BH} = 5\times 10^4 \,  M_{\odot}$) initially exceeds that of the stellar component ($\sim 7\times 10^3\,  M_{\odot}$) of the host galaxy, and the luminosity from BH accretion dominates the starlight. The uncertainties in the IMF of Pop III stars can affect the $M_{\rm BH} / M_{\star}$ ratio derived from our simulation however we do not expect to have an order of magnitude increase in the stellar population. Moreover, we estimated the H$_2$ self-shielding effect by using a local approximation prescription from \citet{1996ApJ...468..269D} whereas N$_{\rm{H_2}}$ is a non-local quantity. This local approximation method has been shown to be accurate within an order of magnitude only \citep{2010MNRAS.402.1249S, 2011MNRAS.418..838W}. This H$_2$ self-shielding factor is valid when only the ground states of H$_2$ are populated. At dense regions where stars form, the H$_2$ LTE would be achieved and 
the shielding would be significantly weaker. This might cause the over estimation of the stellar population in the halo. That being said, the overall effect of having high LW background radiation and using a local H$_2$ self-shielding approximation might balance out each other. Moreover, in our simulation X-ray radiation is regulated by the accretion prescription defined in \ref{sec:acc}. Hence, using different accretion prescription, e.g. Super-Eddington, may further enhance the star formation in the halo. We further stress that, Pop III formation induced by the X-ray radiation is limited to the DCBH scenario and hence we cannot say anything about its statistical significance.

\acknowledgments
AA and JLJ acknowledge support from LANL LDRD Exploratory Research Grant 20170317ER. KSSB was supported by a Porat Postoctoral Fellowship at Stanford University and is supported by XSEDE computing grants TG-AST 190001 AND TG-AST 180052 and the Stampede2 supercomputer at the Texas Advanced Computing Center. JHW was supported by NSF awards AST-1614333 AND OAC1835213 and NASA grant NNX17AG23G. 

\software{\texttt{yt} \citep{2011ApJS..192....9T},  \texttt{Enzo} \citep{2014ApJS..211...19B}}.


\begin{thebibliography}{}
\expandafter\ifx\csname natexlab\endcsname\relax\def\natexlab#1{#1}\fi
\providecommand{\url}[1]{\href{#1}{#1}}

\bibitem[{{Abel} {et~al.}(1997){Abel}, {Anninos}, {Zhang}, \&
  {Norman}}]{1997NewA....2..181A}
{Abel}, T., {Anninos}, P., {Zhang}, Y., \& {Norman}, M.~L. 1997, \na, 2, 181

\bibitem[{{Abel} {et~al.}(2007){Abel}, {Wise}, \&
  {Bryan}}]{2007ApJ...659L..87A}
{Abel}, T., {Wise}, J.~H., \& {Bryan}, G.~L. 2007, \apjl, 659, L87

\bibitem[{{Agarwal} {et~al.}(2013){Agarwal}, {Davis}, {Khochfar}, {Natarajan},
  \& {Dunlop}}]{2013MNRAS.432.3438A}
{Agarwal}, B., {Davis}, A.~J., {Khochfar}, S., {Natarajan}, P., \& {Dunlop},
  J.~S. 2013, \mnras, 432, 3438

\bibitem[{{Agarwal} {et~al.}(2016){Agarwal}, {Smith}, {Glover}, {Natarajan}, \&
  {Khochfar}}]{2016MNRAS.459.4209A}
{Agarwal}, B., {Smith}, B., {Glover}, S., {Natarajan}, P., \& {Khochfar}, S.
  2016, MNRAS, 459, 4209

\bibitem[{{Anninos} {et~al.}(1997){Anninos}, {Zhang}, {Abel}, \&
  {Norman}}]{1997NewA....2..209A}
{Anninos}, P., {Zhang}, Y., {Abel}, T., \& {Norman}, M.~L. 1997, \na, 2, 209

\bibitem[{{Aykutalp} {et~al.}(2013){Aykutalp}, {Wise}, {Meijerink}, \&
  {Spaans}}]{2013ApJ...771...50A}
{Aykutalp}, A., {Wise}, J.~H., {Meijerink}, R., \& {Spaans}, M. 2013, \apj,
  771, 50

\bibitem[{{Aykutalp} {et~al.}(2014){Aykutalp}, {Wise}, {Spaans}, \&
  {Meijerink}}]{2014ApJ...797..139A}
{Aykutalp}, A., {Wise}, J.~H., {Spaans}, M., \& {Meijerink}, R. 2014, \apj,
  797, 139

\bibitem[{{Ba{\~n}ados} {et~al.}(2017){Ba{\~n}ados}, {Venemans},
  {Mazzucchelli}, {Farina}, {Walter}, {Wang}, {Decarli}, {Stern}, {Fan},
  {Davies}, {Hennawi}, {Simcoe}, {Turner}, {Rix}, {Yang}, {Kelson}, {Rudie}, \&
  {Winters}}]{2017arXiv171201860B}
{Ba{\~n}ados}, E., {Venemans}, B.~P., {Mazzucchelli}, C., {et~al.} 2017, ArXiv
  e-prints, arXiv:1712.01860

\bibitem[{Barrow {et~al.}(2018)Barrow, Aykutalp, \& Wise}]{Barrow_2018}
Barrow, K. S.~S., Aykutalp, A., \& Wise, J.~H. 2018, Nature Astronomy, 2,
  987?994.
\newblock \url{http://dx.doi.org/10.1038/s41550-018-0569-y}

\bibitem[{{Begelman} \& {Rees}(1978)}]{1978MNRAS.185..847B}
{Begelman}, M.~C., \& {Rees}, M.~J. 1978, MNRAS, 185, 847

\bibitem[{{Bondi}(1952)}]{1952MNRAS.112..195B}
{Bondi}, H. 1952, MNRAS, 112, 195

\bibitem[{{Bromm} \& {Loeb}(2003)}]{2003ApJ...596...34B}
{Bromm}, V., \& {Loeb}, A. 2003, ApJ, 596, 34

\bibitem[{{Bryan} {et~al.}(2014){Bryan}, {Norman}, {O'Shea}, {Abel}, {Wise},
  {Turk}, {Reynolds}, {Collins}, {Wang}, {Skillman}, {Smith}, {Harkness},
  {Bordner}, {Kim}, {Kuhlen}, {Xu}, {Goldbaum}, {Hummels}, {Kritsuk}, {Tasker},
  {Skory}, {Simpson}, {Hahn}, {Oishi}, {So}, {Zhao}, {Cen}, {Li}, \& {The Enzo
  Collaboration}}]{2014ApJS..211...19B}
{Bryan}, G.~L., {Norman}, M.~L., {O'Shea}, B.~W., {et~al.} 2014, \apjs, 211, 19

\bibitem[{{Draine} \& {Bertoldi}(1996)}]{1996ApJ...468..269D}
{Draine}, B.~T., \& {Bertoldi}, F. 1996, \apj, 468, 269

\bibitem[{{Ebisuzaki} {et~al.}(2001){Ebisuzaki}, {Makino}, {Tsuru}, {Funato},
  {Portegies Zwart}, {Hut}, {McMillan}, {Matsushita}, {Matsumoto}, \&
  {Kawabe}}]{2001ApJ...562L..19E}
{Ebisuzaki}, T., {Makino}, J., {Tsuru}, T.~G., {et~al.} 2001, ApJL, 562, L19

\bibitem[{{Fan} {et~al.}(2006){Fan}, {Strauss}, {Richards}, {Hennawi},
  {Becker}, {White}, {Diamond-Stanic}, {Donley}, {Jiang}, {Kim}, {Vestergaard},
  {Young}, {Gunn}, {Lupton}, {Knapp}, {Schneider}, {Brandt}, {Bahcall},
  {Barentine}, {Brinkmann}, {Brewington}, {Fukugita}, {Harvanek}, {Kleinman},
  {Krzesinski}, {Long}, {Neilsen}, {Nitta}, {Snedden}, \&
  {Voges}}]{2006AJ....131.1203F}
{Fan}, X., {Strauss}, M.~A., {Richards}, G.~T., {et~al.} 2006, \aj, 131, 1203

\bibitem[{{Haehnelt} \& {Rees}(1993)}]{1993MNRAS.263..168H}
{Haehnelt}, M.~G., \& {Rees}, M.~J. 1993, MNRAS, 263, 168

\bibitem[{{Inayoshi} \& {Omukai}(2012)}]{2012AIPC.1480..309I}
{Inayoshi}, K., \& {Omukai}, K. 2012, in American Institute of Physics
  Conference Series, Vol. 1480, American Institute of Physics Conference
  Series, ed. M.~{Umemura} \& K.~{Omukai}, 309--312

\bibitem[{{Johnson} \& {Bromm}(2007)}]{2007MNRAS.374.1557J}
{Johnson}, J.~L., \& {Bromm}, V. 2007, MNRAS, 374, 1557

\bibitem[{{Johnson} \& {Dijkstra}(2017)}]{2017A&A...601A.138J}
{Johnson}, J.~L., \& {Dijkstra}, M. 2017, \aap, 601, A138

\bibitem[{{Kim} {et~al.}(2011){Kim}, {Wise}, {Alvarez}, \&
  {Abel}}]{2011ApJ...738...54K}
{Kim}, J.-h., {Wise}, J.~H., {Alvarez}, M.~A., \& {Abel}, T. 2011, ApJ, 738, 54

\bibitem[{{Kitayama} \& {Yoshida}(2005)}]{2005ApJ...630..675K}
{Kitayama}, T., \& {Yoshida}, N. 2005, \apj, 630, 675

\bibitem[{{Komatsu} {et~al.}(2011){Komatsu}, {Smith}, {Dunkley}, {Bennett},
  {Gold}, {Hinshaw}, {Jarosik}, {Larson}, {Nolta}, {Page}, {Spergel},
  {Halpern}, {Hill}, {Kogut}, {Limon}, {Meyer}, {Odegard}, {Tucker}, {Weiland},
  {Wollack}, \& {Wright}}]{2011ApJS..192...18K}
{Komatsu}, E., {Smith}, K.~M., {Dunkley}, J., {et~al.} 2011, \apjs, 192, 18

\bibitem[{{Kurk} {et~al.}(2007){Kurk}, {Walter}, {Fan}, {Jiang}, {Riechers},
  {Rix}, {Pentericci}, {Strauss}, {Carilli}, \& {Wagner}}]{2007ApJ...669...32K}
{Kurk}, J.~D., {Walter}, F., {Fan}, X., {et~al.} 2007, \apj, 669, 32

\bibitem[{{Latif} {et~al.}(2014){Latif}, {Bovino}, {Van Borm}, {Grassi},
  {Schleicher}, \& {Spaans}}]{2014MNRAS.443.1979L}
{Latif}, M.~A., {Bovino}, S., {Van Borm}, C., {et~al.} 2014, MNRAS, 443, 1979

\bibitem[{{Meijerink} \& {Spaans}(2005)}]{2005A&A...436..397M}
{Meijerink}, R., \& {Spaans}, M. 2005, \aap, 436, 397

\bibitem[{{Mellema} {et~al.}(2006){Mellema}, {Iliev}, {Alvarez}, \&
  {Shapiro}}]{2006NewA...11..374M}
{Mellema}, G., {Iliev}, I.~T., {Alvarez}, M.~A., \& {Shapiro}, P.~R. 2006, New
  Astronomy, 11, 374

\bibitem[{{Mortlock} {et~al.}(2011){Mortlock}, {Warren}, {Venemans}, {Patel},
  {Hewett}, {McMahon}, {Simpson}, {Theuns}, {Gonz{\'a}les-Solares}, {Adamson},
  {Dye}, {Hambly}, {Hirst}, {Irwin}, {Kuiper}, {Lawrence}, \&
  {R{\"o}ttgering}}]{2011Natur.474..616M}
{Mortlock}, D.~J., {Warren}, S.~J., {Venemans}, B.~P., {et~al.} 2011, \nat,
  474, 616

\bibitem[{{Ricotti} {et~al.}(2001){Ricotti}, {Gnedin}, \&
  {Shull}}]{2001ApJ...560..580R}
{Ricotti}, M., {Gnedin}, N.~Y., \& {Shull}, J.~M. 2001, \apj, 560, 580

\bibitem[{{Shang} {et~al.}(2010){Shang}, {Bryan}, \&
  {Haiman}}]{2010MNRAS.402.1249S}
{Shang}, C., {Bryan}, G.~L., \& {Haiman}, Z. 2010, MNRAS, 402, 1249

\bibitem[{{Spaans} \& {Silk}(2006)}]{2006ApJ...652..902S}
{Spaans}, M., \& {Silk}, J. 2006, \apj, 652, 902

\bibitem[{{Umemura} {et~al.}(1993){Umemura}, {Loeb}, \&
  {Turner}}]{1993ApJ...419..459U}
{Umemura}, M., {Loeb}, A., \& {Turner}, E.~L. 1993, ApJ, 419, 459

\bibitem[{{Visbal} {et~al.}(2014){Visbal}, {Haiman}, \&
  {Bryan}}]{2014MNRAS.442L.100V}
{Visbal}, E., {Haiman}, Z., \& {Bryan}, G.~L. 2014, MNRAS, 442, L100

\bibitem[{Volonteri \& Begelman(2010)}]{10.1111/j.1365-2966.2010.17359.x}
Volonteri, M., \& Begelman, M.~C. 2010, \mnras, 409, 1022.
\newblock \url{https://doi.org/10.1111/j.1365-2966.2010.17359.x}

\bibitem[{{Volonteri} {et~al.}(2003){Volonteri}, {Haardt}, \&
  {Madau}}]{2003ApJ...582..559V}
{Volonteri}, M., {Haardt}, F., \& {Madau}, P. 2003, ApJ, 582, 559

\bibitem[{{Volonteri} \& {Rees}(2005)}]{2005ApJ...633..624V}
{Volonteri}, M., \& {Rees}, M.~J. 2005, ApJ, 633, 624

\bibitem[{{Wise} \& {Abel}(2008)}]{2008ApJ...685...40W}
{Wise}, J.~H., \& {Abel}, T. 2008, \apj, 685, 40

\bibitem[{{Wise} \& {Abel}(2011)}]{Wise11_Moray}
---. 2011, MNRAS, 414, 3458

\bibitem[{Wise {et~al.}(2019)Wise, Regan, O?Shea, Norman, Downes, \&
  Xu}]{Wise_2019}
Wise, J.~H., Regan, J.~A., O?Shea, B.~W., {et~al.} 2019, Nature, 566,
  85?88.
\newblock \url{http://dx.doi.org/10.1038/s41586-019-0873-4}

\bibitem[{{Wise} {et~al.}(2008){Wise}, {Turk}, \& {Abel}}]{2008ApJ...682..745W}
{Wise}, J.~H., {Turk}, M.~J., \& {Abel}, T. 2008, ApJ, 682, 745

\bibitem[{{Wise} {et~al.}(2012){Wise}, {Turk}, {Norman}, \&
  {Abel}}]{2012ApJ...745...50W}
{Wise}, J.~H., {Turk}, M.~J., {Norman}, M.~L., \& {Abel}, T. 2012, \apj, 745,
  50

\bibitem[{{Wolcott-Green} \& {Haiman}(2012)}]{2012MNRAS.425L..51W}
{Wolcott-Green}, J., \& {Haiman}, Z. 2012, \mnras, 425, L51

\bibitem[{{Wolcott-Green} {et~al.}(2011){Wolcott-Green}, {Haiman}, \&
  {Bryan}}]{2011MNRAS.418..838W}
{Wolcott-Green}, J., {Haiman}, Z., \& {Bryan}, G.~L. 2011, \mnras, 418, 838

\bibitem[{{Wolcott-Green} {et~al.}(2020){Wolcott-Green}, {Haiman}, \&
  {Bryan}}]{2020arXiv200105498W}
---. 2020, arXiv e-prints, arXiv:2001.05498

\bibitem[{{Woods} {et~al.}(2017){Woods}, {Heger}, {Whalen}, {Haemmerl{\'e}}, \&
  {Klessen}}]{2017ApJ...842L...6W}
{Woods}, T.~E., {Heger}, A., {Whalen}, D.~J., {Haemmerl{\'e}}, L., \&
  {Klessen}, R.~S. 2017, \apjl, 842, L6

\bibitem[{{Wu} {et~al.}(2015){Wu}, {Wang}, {Fan}, {Yi}, {Zuo}, {Bian}, {Jiang},
  {McGreer}, {Wang}, {Yang}, {Yang}, {Thompson}, \&
  {Beletsky}}]{2015Natur.518..512W}
{Wu}, X.-B., {Wang}, F., {Fan}, X., {et~al.} 2015, Nature, 518, 512

\end{thebibliography}
\end{document}